\documentclass[aps,prb,twocolumn,a4paper,10pt,showkeys,showpacs,preprintnumbers,floatfix,superscriptaddress]{revtex4-1}

\usepackage{graphicx}	
\usepackage{dcolumn}		% Align table columns on decimal point
\usepackage{bm}				% bold math
\usepackage{float}
\usepackage{amsmath}
\usepackage{amssymb}
\usepackage{amsfonts}
\usepackage{times}
\usepackage{tabularx}
\usepackage{xcolor,colortbl}
\usepackage{mathrsfs}
\usepackage{amsbsy}

\begin{document}
\title{Controlled anisotropic dynamics of tightly bound skyrmions in a synthetic ferrimagnet due to skyrmion-deformation mediated by induced uniaxial in-plane anisotropy}
\author{P. E. Roy}
\email{per24.ac.uk}
\affiliation{Hitachi Cambridge Laboratory, J. J. Thomson Avenue, CB3 OHE, Cambridge, United Kingdom}
\author{Rub\'{e}n M. Otxoa}
\affiliation{Hitachi Cambridge Laboratory, J. J. Thomson Avenue, CB3 OHE, Cambridge, United Kingdom}
\affiliation{ Donostia International Physics Center, Paseo Manuel de Lardizabal 4, Donostia-San Sebastian 20018, Spain}
\author{C. Moutafis}
\affiliation{School of Computer Science, University of Manchester, Manchester M13 9PL, United Kingdom}
%\thanks{}
\date{\today}
\begin{abstract}
We study speed and skew deflection-angle dependence on skyrmion deformations of a tightly bound two-skyrmion state in a synthetic ferrimagnet. We condsider here, an in-plane uniaxial magnetocrystalline anisotropy-term in order to induce lateral shape distortions and an overall size modulation of the skyrmions due to a reduction of the effective out-of-plane anisotropy, thus affecting the skyrmion speed, skew-deflection and inducing anisotropy in these quantities with respect to the driving current-angle. Because of frustrated dipolar interactions in a synthetic ferrimagnet, sizeable skyrmion deformations can be induced with relatively small induced anisotropy constants and thus a wide range of tuneability can be achieved. We also show analytically, that a consequence of the skyrmion deformation can, under certain conditions cause a skyrmion deflection with respect to driving-current angles, unrelated to the topological charge. Results are analyzed by a combination of micromagnetic simulations and a compound particle description within the Thiele-formalism from which an over-all mobility tensor is constructed. This work offers an additional path towards in-situ tuning of skyrmion dynamics.
\end{abstract}
\maketitle

\section{Introduction}
Magnetic skyrmions have been extolled as candidates for the constituent information carriers in spintronic devices such as racetrack memories and logic circuits \cite{Racetrack1,Racetrack2,Racetrack3,Racetrack4,Racetrack5,logic1}. Their advantage over conventional domain walls are that they are less sensitive to edge defects and can be driven at much lower current densities. However, similar to the Hall-effect for an electrically charged particle, a magnetic skyrmion viewed as a quasi-particle endowed with a topological charge, exhibits a deflection known as the skyrmion Hall-effect \cite{SkyHa1,SkyHa2}, resulting in a skew deflection known as skyrmion Hall-angle, $\Theta_{Sk}$. This can lead to detrimental annihilation at device-boundaries at high enough driving amplitudes. In order to overcome this, there have been proposals suggesting the usage of both intrinsic \cite{AFintr1,AFintr2,AFintr3} and synthetic antiferromagnets \cite{SAFSky0,SAFSky} (SAFs) with identical but oppositely magnetized sublattices, whereby the direction of the gyroforce on a skyrmion in one sublattice is equal but opposite in direction for the skyrmion on the other sublattice \cite{AFintr1}. This is because, by virtue of satisfying the antiferromagnetic coupling they have opposite topological charge. The two deflection forces then fully cancel out and the compound object moves without deflection. An advantage of antiferromagnetic systems is the robustness against external field perturbations. However this means that the manipulation and detection of antiferromagnetic textures is generally a difficult task. Devices based on single layer ferromagnets as the functional component have the advantage of easy detection schemes but are senstitive to external fields. However in ferrimagnetic systems, a low net moment is present, offering reasonable detectibility and at the same time a good degree of robustness against disturbing stray-fields. A net moment though, means a finite skyrmion-Hall angle (even if it is much reduced compared to a single layer ferromagnetic system). An additional benefit of synthetic systems is the large tuneability of material properties, but the conclusions extolled herein should be valid also for instrinsic ferrimagnets. Apart from the desire to achieve some degree of control over $\Theta_{Sk}$, another important dynamical property to tune and/or enhance is the speed of skyrmion propagation. Enhancement is desired due to increased operating frequency of a device. Tuning in general would find usage wherever the skyrmionic device is a subset of a bigger multifunctional device, whereby timescales need to be matched at different points of device functionality in space. In order to modulate the dynamical properties of a skyrmionic magnetic device, one may consider tailoring of material properties during fabrication and/or affecting the ready-made device by external means during operation. We refer to the former as intrinsic and the latter as extrinsic tuning, respectively. In particular, extrinsic tuning offers manipulation on the fly. Different approaches have previously been presented for dynamical control such as: by mismatching the saturation magnetization of the constituent ferromagnetic (FM) layers in an SFIM \cite{modulation1}, spatially uniform modulation of the perpendicular anisotropy \cite{modulation2}, perpendicular anisotropy gradients \cite{modulation3,modulation4} and radial magnetic field gradients \cite{modulation5}. In such a scenario (for a given topological charge) the skyrmion radius is isotropically varied and thus its dynamical behaviour is isotropic in the plane of propagation with respect to the driving current-angle. There have recently been some very interesting works considering the effect of inducing anisotropic Dzyaloshinskii-Moriya interaction in single layer FMs as a control-knob for skyrmion dynamics (speed and skew-angle), offering also due to skyrmion-shape distortion anisotropic dynamical behaviour with respect to the direction of the driving excitation mechanism, \cite{modulation6,modulation7}. The anisotropic dynamics adds an additional degree of freedom in tuning the skyrmion dynamical behaviour.

In this work, we focus on synthetic ferrimagnets (SFIMs) for reasons mentioned in the preceeding paragraph and consider here tuning the dynamics due to deformation of a bound two-skyrmion texture via only an induced uniaxial in-plane anisotropy term, achievable by e.g. an inverse magnetostrictive effect whereby a mechanical uniaxial stress induces a uniaxial contribution to the magnetocrystalline energy \cite{Bozorth}. The driving mechanism is provided by spin-Hall-effect induced torques. By utilizing a synthetic ferrimagnet, we are able to induce large skyrmion deformations for relatively small values of the induced uniaxial in-plane anisotropy constants and predict a wide tuneability of the bound skyrmion skew-deflection, speed and degree of anisotropy of the said quantities with respect to the driving current-angle. In addition, it is shown, that, given a deviation from circularly shaped skyrmions, that for driving current angles different from multiples of $\pi/2$, there is a contribution to the skyrmion deflection away from the driving-current direction which is independent of the topological charge.
\section{Methods}

\subsection{Thiele approach and the effective skyrmion mobility tensor}
\begin{figure}
\includegraphics[width=\columnwidth]{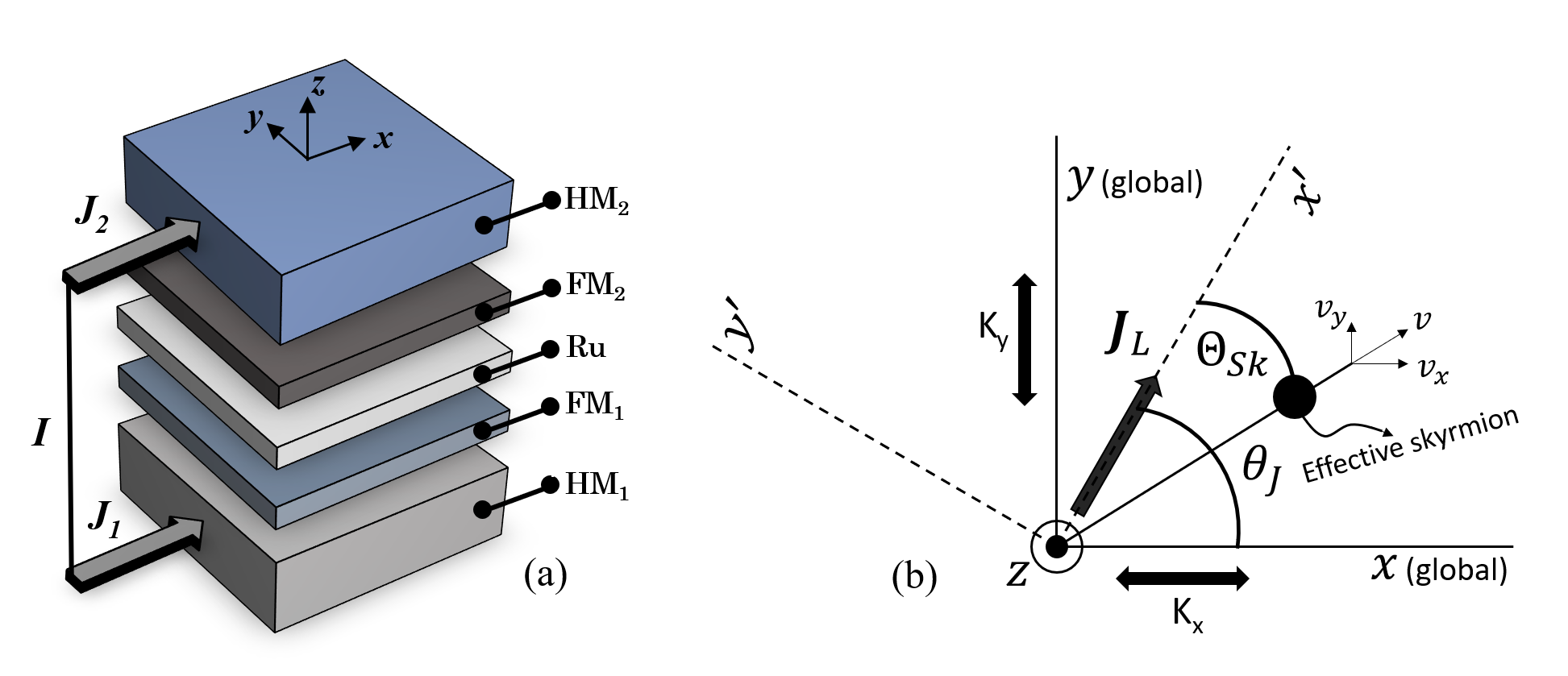}\caption{\label{fig:Figure1} (a) Geometry and constituent layers of the stack considered in this work. (b): Top view of the global ($xy$) and primed/local ($x'y'$) coordinate systems. Here $J_{L}$ is the current density in the $L$:th FM layer whose direction is at an angle $\theta_{J}$ with respect to the global system. $\Theta_{Sk}$ of the bound skyrmion is defined as the angle of deviation from the direction of the current density. Double arrows signify the considered directions of induced uniaxial anisotropies (with anisotropy constants $K_{x}$ and $K_{y}$ depending on whether the easy direction is induced along $x$ or $y$, respectively). The velocity components of the effective skyrmion in the global coordinate system are designated $v_{x}$ and $v_{y}$, whereas the speed is denoted as $v$.}
\end{figure}

The system under consideration is schematically depicted in Fig. (\ref{fig:Figure1}), whereby two FM layers ($\text{FM}_{1}$ and $\text{FM}_{2}$) are separated by a Ru spacer, magnetically coupled via antiferromagnetic (RKKY-type) and dipolar interactions. Each FM layer is also coupled to a heavy metal (HM) which promotes interfacial Dzyaloshinskii-Moriya (IDMI)-interaction. Each FM layer contains one skyrmion. The two skyrmions are considered to be strongly antiferromagnetically coupled, i.e. tightly bound such that they move as one compound unit without internal nor relative dynamics. The means of propulsion of this compound skyrmion is supplied by torques exerted by a spin-accummulation due to the Spin-Hall-Effect (SHE). The SHE is produced by passing a current $\mathbf{I}$ through the HMs, generating current-densities $\mathbf{J}_{1}$ and $\mathbf{J}_{2}$ in $\text{HM}_{1}$ and $\text{HM}_{2}$, respectively (see Fig. \ref{fig:Figure1} (a)). We consider an arbitrary in-plane current-direction. Thus we define a global coordinate system and a primed/local system, whereby the current direction defines the rotation of primed coordinates with respect to global coordinates, as shown in Fig. (\ref{fig:Figure1} (b)).

In order to analytically describe the dynamics of the skyrmions, we turn to the Thiele equation \cite{Thiele2} with Spin-Hall forces \cite{Racetrack2,SkyHa1,SkyHa2}. We consider the tightly bound skyrmions in dynamic equilibrium and sum the forces to zero according to \cite{modulation1}:
\begin{equation} \sum_{L}\frac{\mu_{0}d_{L}M_{L}}{\gamma_{L}}\left\lbrace-Q_{L}\left[\hat{\mathbf{z}}\times\mathbf{v}\right]-\alpha_{L}\pmb{\mathsf{D}}_{L}\cdot\mathbf{v}+4\pi b'_{L}\pmb{\mathsf{I}}_{L}\cdot\mathbf{J}_{L}\right\rbrace=\mathbf{0}.\label{eq:forcebalance}
\end{equation}
The subscript $L=1,2$ denotes the FM layer number (Fig. \ref{fig:Figure1} (a)), $\gamma_{L}$ is the gyromagnetic ratio, $\mu_{0}$, the permeability in vacuum, $M_{L}$ the saturation magnetization and $d_{L}$ the layer thickness. The first term in Eq. (\ref{eq:forcebalance}) represents the gyroforce with topological charge \cite{Papanicolaou} $Q_{L}=\frac{1}{4\pi}\iint\hat{\mathbf{m}}_{L}\cdot\left[\partial_{x}\hat{\mathbf{m}}_{L}\times\partial_{y}\hat{\mathbf{m}}_{L}\right]dxdy$ (the topological charge being a component of the gyrovector $\mathbf{G}_{L}$ through the relation $\mathbf{G}_{L}=\left[0\text{ } 0\text{ } -4\pi Q_{L}\right]$), with $\hat{\mathbf{m}}_{L}=\mathbf{M}_{L}/M_{L}$. The second term constitutes a dissipative drag force with an associated dissipation tensor  $\pmb{\mathsf{D}}_{L}=\text{4}\pi$$\left[\begin{smallmatrix} 
\left(\mathsf{D}_{ii}\right)_{L} & \left(\mathsf{D}_{ij}\right)_{L} \\
\left(\mathsf{D}_{ji}\right)_{L} & \left(\mathsf{D}_{jj}\right)_{L} 
\end{smallmatrix}\right]$, whose elements are given by $\left(\mathsf{D}_{ij}\right)_{L}=\frac{1}{4\pi}\iint\left[\partial_{i}\hat{\textbf{m}}_{L}\cdot\partial_{j}\hat{\textbf{m}}_{L}\right]dxdy$ \cite{Thiele2,Racetrack2}.  The last term is the force exerted by the SHE-induced torque with $b'_{L}=\frac{\gamma_{L}\hbar\phi_{L}o_{L}}{\mu_{0}2\left|e\right|M_{L}d_{L}}$ \cite{SHEamplitude} and $\left[I_{qr}\right]_{L}=\frac{-1}{4\pi}\iint\left[\partial_{q}\hat{\textbf{m}}_{L}\times\hat{\textbf{m}}_{L}\right]_{s}\epsilon_{sr}\text{d}x\text{d}y$ \cite{Racetrack2,Btensor}; $q,r\in \lbrace x,y\rbrace$ and $\epsilon_{sr}$ is the Levi-Civita symbol. Further, $\hbar$ is Planck's reduced constant, $\phi_{L}$ the intrinsic spin-Hall angle and $e$ is an electron's charge. The unit magnitude factor $o_{L}$ takes into account the effect of HM/FM stacking order, on the direction of the resulting spin-accummulation. As $\text{HM}_{1}$ is under $\text{FM}_{1}$ and $\text{HM}_{2}$ is above $\text{FM}_{2}$ $o_{1}$ and $o_{2}$ have opposite sign. We set here, $o_{1}=1$ and $o_{2}=-1$. For brevity we assign a tensor $\pmb{\mathsf{S}}_{L}=b'_{L}\pmb{\mathsf{I}}_{L}$. Note that the units of $\pmb{\mathsf{S}}$ is [$\text{m}^{3}$/As]  (velocity per unit current density) and as such may be viewed as a SHE-related mobility tensor in the absence of other forces. We reiterate, that the underlying assumptions in stating the problem by Eq. (\ref{eq:forcebalance}), are that the internal structures of the skyrmions are rigid and that the antiferromagnetic coupling between them is strong enough such that they move together without any relative dynamics. Now, intrinsic tuning of $\Theta_{Sk}$ and speed of the tightly bound system would be to consider a mismatch of the saturation magnetization, gyromagnetic ratio or thickness between the two FM layers. This is more easily seen by rewriting Eq. (\ref{eq:forcebalance}) as $-Q^{e}\left[\hat{\mathbf{z}}\times\mathbf{v}\right]-\pmb{\mathsf{D}}^{e}\cdot\mathbf{v}+w\pmb{\mathsf{S}}_{1}\cdot\mathbf{J}_{1}+\pmb{\mathsf{S}}_{2}\cdot\mathbf{J}_{2}=\mathbf{0}$, where $Q^{e}=wQ_{1}+Q_{2}$ and $\pmb{\mathsf{D}}^{e}=w\alpha_{1}\pmb{\mathsf{D}}_{1}+\alpha_{2}\pmb{\mathsf{D}}_{2}$, with $w$=$\frac{\gamma_{2}d_{1}M_{1}}{\gamma_{1}d_{2}M_{2}}$. Further, we write the current density $\mathbf{J}_{2}$ in terms of $\mathbf{J}_{1}$ and make the reasonable assumption that $\mathbf{J}_{1}$ and $\mathbf{J}_{2}$ point in the same direction such that $\mathbf{J}_{2}=k_{s}\mathbf{J}_{1}$, where $k_{s}$ is a real scaling factor. Thus the inclusion of different magnitudes of current densities in the top and bottom FMs is retained. Now, we can define; $\pmb{\mathsf{S}}^{e}=w\pmb{\mathsf{S}}_{1}+k_{s}\pmb{\mathsf{S}}_{2}$. Therefore, here, the SHE-mobility of the compound particle is a function of the ratio of current-density amplitudes in the FM layers. The Thiele equation for the single compound particle reads: 
\begin{equation}
-Q^{e}\left[\hat{\mathbf{z}}\times{\mathbf{v}}\right]-\pmb{\mathsf{D}}^{e}\cdot\mathbf{v}+\pmb{\mathsf{S}}^{e}\cdot\mathbf{J}_{1}=\mathbf{0}. \label{eq:effective}
\end{equation}
Viewing the system in terms of this compound particle description with effective properties, which henceforth is called effective skyrmion, it is easily seen that we can modulate $Q^{e}$, $\pmb{\mathsf{D}}^{e}$ and $\pmb{\mathsf{S}}^{e}$ by varying $w$ and thus affect the dynamical properties such as speed and $\Theta_{Sk}$ of the effective skyrmion. However, a path towards \textit{in-situ} modulation of the skyrmion dynamics is rather to utilize the fact that the elements of the tensors in Eq. (\ref{eq:forcebalance}) and thus in Eq. (\ref{eq:effective}) are determined by the geometry of the texture \cite{modulation2}. Provided a suitable material system is available, one route towards \textit{in-situ} manipulation is by an induced in-plane anisotropy in order to deform the magnetic texture (induced by e.g. the inverse magnetostrictive effect, whereby an imposed mechanical stress induces a magnetic uniaxial anisotropy). For the type of skyrmion deformation considered here (expansions along principal axes), we assume $\mathsf{D}^{e}_{xy}=\mathsf{D}^{e}_{yx}=0$ and $\mathsf{S}^{e}_{xy}=\mathsf{S}^{e}_{yx}=0$ (which was verified when evaluating the tensor components from micromagnetically obtained magnetization distributions). From Eq. (\ref{eq:effective}) the velocity components are then $v_{x}=\frac{Q^{e}\mathsf{S}_{yy}J_{1y}+\mathsf{D}^{e}_{yy}\mathsf{S}_{xx}J_{1x}}{\left(Q^{e}\right)^{2}+\mathsf{D}^{e}_{xx}\mathsf{D}^{e}_{yy}}$ and $v_{y}=\frac{\mathsf{D}^{e}_{xx}\mathsf{S}_{yy}J_{1y}-Q^{e}\mathsf{S}_{xx}J_{1x}}{\left(Q^{e}\right)^{2}+\mathsf{D}^{e}_{xx}\mathsf{D}^{e}_{yy}}$. Thus we can write the velocity of the effective skyrmion in terms of an over-all effective mobility tensor $\pmb{\mathsf{\mu}}^{e}$. Remembering that the current density in the first FM is to be used and setting $\mathbf{J}=\mathbf{J}_{1}$, we may write:
\begin{align}
v_{i}=\mu_{ij}^{e}J_{j} \label{eq:speeds}\\
\pmb{\mu}^{e}=
\begin{bmatrix}
\frac{\mathsf{D}_{yy}^{e}\mathsf{S}_{xx}^{e}}{\left(Q^{e}\right)^{2}+\mathsf{D}_{xx}^{e}\mathsf{D}_{yy}^{e}} & \frac{Q^{e}\mathsf{S}_{yy}^{e}}{\left(Q^{e}\right)^{2}+\mathsf{D}_{xx}^{e}\mathsf{D}_{yy}^{e}}\\ \frac{-Q^{e}\mathsf{S}_{xx}^{e}}{\left(Q^{e}\right)^{2}+\mathsf{D}_{xx}^{e}\mathsf{D}_{yy}^{e}} & \frac{\mathsf{D}_{xx}^{e}\mathsf{S}_{yy}^{e}}{\left(Q^{e}\right)^{2}+\mathsf{D}_{xx}^{e}\mathsf{D}_{yy}^{e}} \label{eq:mobility}
\end{bmatrix}
\end{align}
Eqs. (\ref{eq:speeds}-\ref{eq:mobility}) constitute the velocity components of the compound particle with respect to the global coordinate system (see Fig. \ref{fig:Figure1}(b)) and as such, determining the $\Theta_{Sk}$ from the ratio $v_{y}/v_{x}$ would give an angle with respect to the global $x$-axis. Thus in order to study the $\Theta_{Sk}$-dependence on current direction we must define it with respect to a rotated coordinate system defined by the current direction. If we denote the angle of the uniform current density with respect to the global $x$-axis by $\theta_{J}$, and $\mathbf{J}=J\left[\cos\left(\theta_{J}\right)\text{ }\sin\left(\theta_{J}\right)\right]^{\text{T}}$, then we can express $v_{x}$ and $v_{y}$ in terms of the primed coordinate system as $v_{x'}=\text{cos}\left(\theta_{J}\right)v_{x}+\text{sin}\left(\theta_{J}\right)v_{y}$ and $v_{y'}=-\text{sin}\left(\theta_{J}\right)v_{x}+\text{cos}\left(\theta_{J}\right)v_{y}$. Consequently, $\Theta_{\text{Sk}}=\text{atan}\left(\frac{v_{y'}}{v_{x'}}\right)$.
\subsection{Micromagnetic technique}
In order to calculate the mobility tensor under various states with different induced in-plane anisotropies, we use equilibrium configurations determined by micromagnetic simulations. To stabilize the antiferromagnetic alignment between a skyrmion in $\text{FM}_{1}$ and another skyrmion in $\text{FM}_{2}$, it is necessary for them to have the same handedness \cite{SAFSky}. The handedness is determined by the sign of the IDMI. The sign itself, depends on; i) intrinsic material properties and ii) whether the HM is coupled below or above a FM, since opposite stacking order will change the sign of the IDMI \cite{DMIsign1,DMIsign2}. For the former, we denote the intrinsic IDMI strengths by $D_{L}$ for a given layer $L$. The latter contribution is taken into account for by a multiplicative factor $f_{L}$ where we set $f_{1}=+1$ and $f_{2}=-1$. When $f_{1}D_{1}$=$f_{2}D_{2}$ then textures in both FMs will have the same handedness \cite{SAFSky}. Thus for the stack considered here, HMs need to be chosen such that $D_{2}$=$-D_{1}$ \cite{SAFSky}. Furthermore, in order for the AFM coupled skyrmions to move in the same direction when current is passed through both HMs, the intrinsic Spin-Hall angles, $\phi_{L}$ of the two HMs need to be opposite in sign ($\phi_{2}$=$-\phi_{1}$) \cite{SAFSky}. One possible HM material pair could be e.g. Pt and W \cite{SAFSky}. The total energy density considered of a given FM layer $L$ is:
\begin{equation}
\label{Total_Energy}
\begin{aligned}
\epsilon_{L}=A_{L}\sum\limits_{i=1}^{3}\left|\nabla m_{i,L}\right|^{2}+\frac{\sigma}{t_{\text{Ru}}}\left(1-\hat{\textbf{m}}_{L=1}\cdot \hat{\textbf{m}}_{L=2}\right)
\\
+f_{L}D_{L}\left[m_{z,L}\left(\nabla\cdot
\hat{\textbf{m}}_{L} \right)-\left(\hat{\textbf{m}}_{L}\cdot\nabla\right)m_{z,L}\right]
\\
+K_{U,L}^{\perp}\left(1-\left(\hat{\textbf{m}}_{L}\cdot \hat{\textbf{z}}\right)^{2}\right)+K_{U,L}^{\parallel}\left(1-\left(\hat{\textbf{m}}_{L}\cdot \hat{\textbf{u}}_{\parallel}\right)^{2}\right)\\
-\frac{1}{2}\mu_{0}M_{L}\hat{\textbf{m}}_{L}\cdot\textbf{H}_{m,L}.
\end{aligned}
\vspace{5mm}
\end{equation}
In Eq. (\ref{Total_Energy}), $A_{L}$ are the intra-layer exchange stiffness constants, $\sigma$ the inter-layer exchange coupling constant (here antiferromagnetic) between the layers, $t_{\text{Ru}}$ is the thickness of the Ru spacer, $K^{\perp}_{U,L}$ are out-of-plane uniaxial magnetocrystalline anisotropy constants, $K^{\parallel}_{U,L}$ are in-plane uniaxial magnetocrystalline anisotropy constants with easy directions $\hat{\textbf{u}}_{\parallel}$. We shall henceforth use the denomination $K_{x}$ if $\hat{\textbf{u}}_{\parallel}=\hat{\textbf{x}}$ to describe induced in-plane anisotropy along the $x$-direction and $K_{y}$ if $\hat{\textbf{u}}_{\parallel}=\hat{\textbf{y}}$ for the $y$-direction. It is also to be understood that in any state of induced anisotropy we consider, $K^{\parallel}_{U,1}=K^{\parallel}_{U,2}$. Further, $\textbf{H}_{m,L}$ is the magnetostatic field, is evaluated in the the entire computational domain. The interaction terms in Eq. (\ref{Total_Energy}) of the main text give rise to an effective field $\textbf{H}_{\text{eff},L}$=$\frac{-1}{\mu_{0}M_{L}}\frac{\partial\epsilon_{L}}{\partial\hat{\textbf{m}}_{L}}$ acting on the magnetization. In particular the IDMI field is $\textbf{H}_{\text{IDMI},L}$=$-\frac{2f_{L}D_{L}}{\mu_{0}M_{L}}\left[\left( \nabla\cdot \hat{\textbf{m}}_{L}\right)\hat{\textbf{z}}-\nabla m_{z,L}\right]$ \cite{Martineztech,ThiavilleBC} and the inter-layer exchange field at a site $\textit{i}$ due to interaction with a site $\textit{j}$ is $\textbf{H}_{\text{RKKY},\textit{i}}=\frac{2\sigma}{\text{t}_{\text{Ru}}\mu_{0}M_{i}}\hat{\textbf{m}}_{j}$. We consider here that this RKKY-like interaction occurs between the computational cells separated purely along the $\hat{\textbf{z}}$-axis. Due to the IDMI, Neumann boundary conditions on the lateral surfaces of the FMs are $\frac{\partial \hat{\textbf{m}}_{L}}{\partial \hat{\textbf{n}}_{L}}=-\frac{f_{L}D_{1,2}}{2A_{L}}\left(\hat{\textbf{z}}\times\hat{\textbf{n}}_{L}\right)\times \hat{\textbf{m}}_{L}$, where $\hat{\textbf{n}}_{L}$ is the unit normal vector to the lateral surfaces \cite{Martineztech,ThiavilleBC}. For surfaces normal to the plane, homogeneous Neumann conditions are used. The magnetostatic field, $\textbf{H}_{m,L}$ is computed as the spatial convolution between a demagnetizing tensor \cite{Newell} and the magnetization distributions by Fast Fourier Transform techniques \cite{Bagneres}. In the magnetostatic field computation, for the near-field, we use analytical formulae for the demagnetizing tensor by Newell and Dunlop \cite{Newell} and for the far field, at relative cell distances larger than 40, the point dipole approximation is used. The dynamics of the system is modelled by solving the Landau-Lifshitz-Gilbert (LLG) equation, with added Spin-Hall-Effect torques \cite{Martineztech}:
\begin{equation}
\label{eq:LLG}
\begin{aligned}
\frac{\partial\hat{\textbf{m}}_{L}}{\partial t}=-\gamma_{L}\hat{\textbf{m}}_{L}\times\textbf{H}_{\text{eff},L}+\alpha_{L}\hat{\textbf{m}}_{L}\times \frac{\partial\hat{\textbf{m}}_{L}}{\partial t}
\\
-\left|b'_{L} J_{L}\right|\left[\hat{\textbf{m}}_{L}\times\left(\hat{\textbf{m}}_{L}\times\hat{\textbf{p}}_{L}\right)\right],
\end{aligned}
\end{equation}
where $\gamma_{L}$=2.21$\times\text{10}^{\text{5}}$m/As is the gyromagnetic ratio, $\alpha_{L}$ is the Gilbert damping, $b'_{L}=\frac{\hbar\gamma_{L}\phi_{L}}{\mu_{0}\text{2} e M_{L}d_{L}}$ \cite{SHEamplitude} with $\phi_{L}$ being the intrinsic Spin-Hall angles, $e$ the electron charge, $d_{\text{L}}$ are the FM layers' thicknesses, $J_{L}$ are the current density amplitudes and $\hat{\textbf{p}}_{L}$ are the directions of the spin-accumulation acting on layer $L$ due to the SHE at the FM/HM interfaces. The spin-accumulation due to a current density along direction $\hat{\textbf{j}}_{L}$ (unit vector) present in $\text{HM}_{L}$ is
$\hat{\textbf{p}}_{L}=\text{sgn}\phi_{L}\left(\hat{\textbf{j}}_{L}\times\hat{\textbf{n}}_{\text{HM-FM},L}\right)$, where $\hat{\textbf{n}}_{\text{HM-FM},L}$ is the unit normal vector directed from a HM towards a FM layer\cite{Martineztech}. For computational simplicity, Eq. (\ref{eq:LLG}) is cast into explicit form and solved by a fifth order Runge-Kutta integration scheme \cite{Numrec}. The lateral space considered is a rectangular domain of dimensions $\text{1200}\times\text{768}$ n$\text{m}^{2}$ whereas the comprising FM layers and spacer layer all have a thickness of 0.8 nm. The domain is discretized into $\text{1.5}\times\text{1.5}\times\text{0.8}$ n$\text{m}^{3}$ cells. Material parameters considered are the following: $A_{1}=A_{2}=\text{20}$ p$\text{J}$/m, $M_{1}=\text{0.6}$ MA/m, $M_{2}=\text{0.75}$ MA/m, $K^{\perp}_{U,1}=K^{\perp}_{U,2}=$ 0.6 MJ/$\text{m}^{3}$, $\gamma_{1}=\gamma_{2}=$ 2.21$\times \text{10}^{5}$ m/As, $\alpha_{1}=\alpha_{2}=$ 0.1, $D_{1}=$ 2.8 mJ/$\text{m}^{2}$, $D_{1}=-$ 2.5 mJ/$\text{m}^{2}$, $\phi_{1}=$ 0.1, $\phi_{2}=-\text{0.1}$, $\sigma=-\text{0.5}$ mJ/$\text{m}^{2}$ and $K^{\parallel}_{U,1}=K^{\parallel}_{U,2}$ is varied in the range [0,0.09] MJ/$\text{m}^{3}$. 
\begin{figure*}[t]
\includegraphics[width=\textwidth]{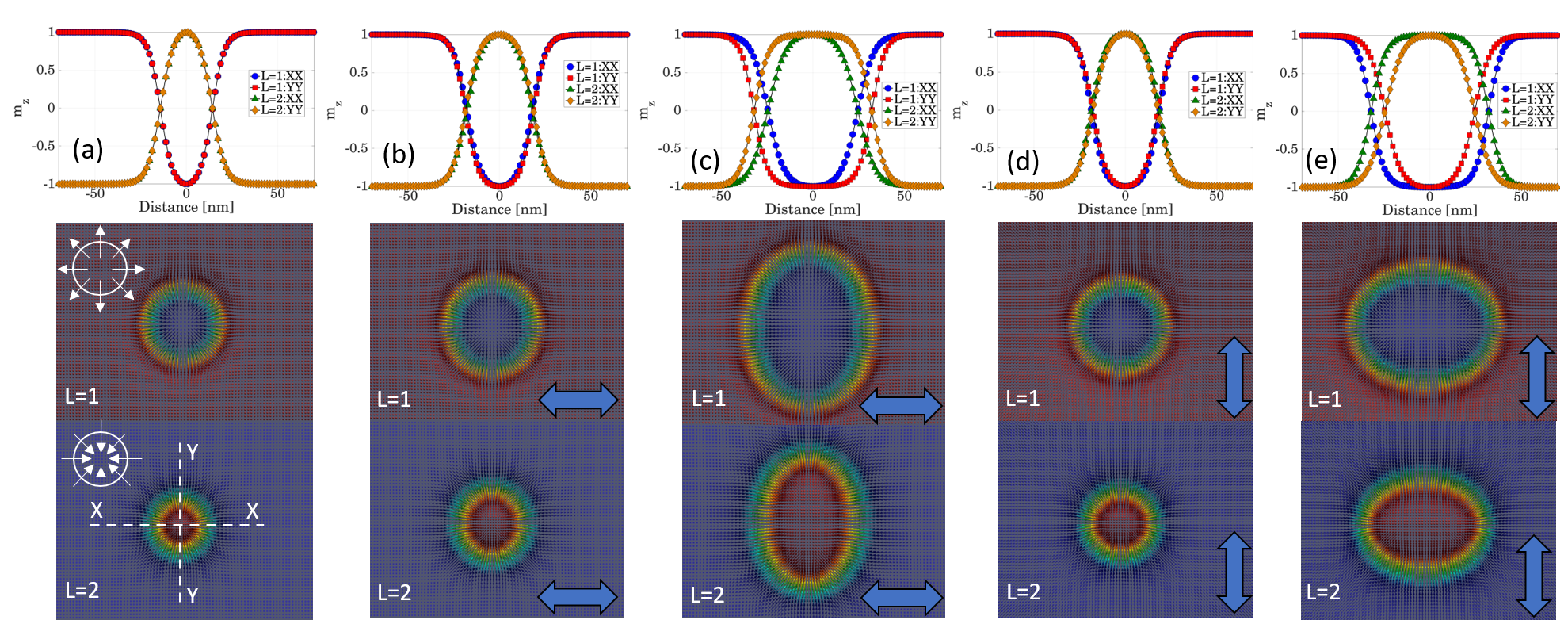}\caption{\label{fig:Figure2} (Colour online) Skyrmion $m_{z}$-profiles along two cuts, XX and YY for various states of induced $K^{\parallel}_{U,L}$. Below each profile is shown the corresponding vector-plots of the magnetization distributions in the bottom (L=1) and top (L=2) layers with colorcoding corresponding to $m_{z}$; red = +1 and dark blue= -1. The circles with arrows in the magnetization distribution corresponding to $K_{x}=K_{y}=0$ indicate the in-plane components of the skyrmions. The direction of induced anisotropy is indicated by thick double arrows in the vector-plots : (a): $K_{x}=K_{y}=$ 0. (b): $K_{x}=$ 0.05 MJ/$\text{m}^{3}$. (c): $K_{x}=$ 0.09 MJ/$\text{m}^{3}$. (d): $K_{y}=$ 0.05 MJ/$\text{m}^{3}$. (e): $K_{y}=$ 0.09 MJ/$\text{m}^{3}$.}
\end{figure*}
For determining static equilibrium configurations, initial conditions close to that of two antiferromagnetically coupled skyrmions were imposed and the system let to freely ring down until a convergence was reached in the whole computational domain (each layer satisfying $\frac{1}{M_{L}}\left|\hat{\mathbf{m}}_{L}\times\mathbf{H}_{\text{eff},L}\right|\le \text{10}^{-6}$).
In the calculation of $Q^{e}$, $\mathsf{D}^{e}$ and $\mathsf{S}^{e}$ from the static equilibrium distributions, contributions from magnetization-canting at and close to the boundaries were excluded by removing 45 computational cells into the computational domain along all lateral normals. For all current-driven dynamical calculation we consider $J_{1}\ne J_{2}$, specifically $J_{1}=\text{10}^{11}$ and $J_{2}=2\times\text{10}^{11}$  A/$\text{m}^{2}$ ( i.e. $k_{s}=2$). For the evaluation of $\Theta_{Sk}$ from the micromagnetic simulations, skyrmion positions versus time in each FM layer were tracked by the moments of topological density \cite{Papanicolaou}. Simulations for each case of induced anisotropy were run until steady state motion of the skyrmion pair was achieved. Care was taken to only extract data for positions sufficiently far away from the boundaries in order to exclude repulsive boundary-interactions. From the extracted longitudinal and transverse speeds, $v_{x}$ and $v_{y}$ the speed $v$ was obtained and $\Theta_{Sk}$ was extracted from the velocities expressed in the primed coordinate system according to the preceding section.
\section{Results and Discussion}
\begin{figure}[H]
\includegraphics[width=\columnwidth]{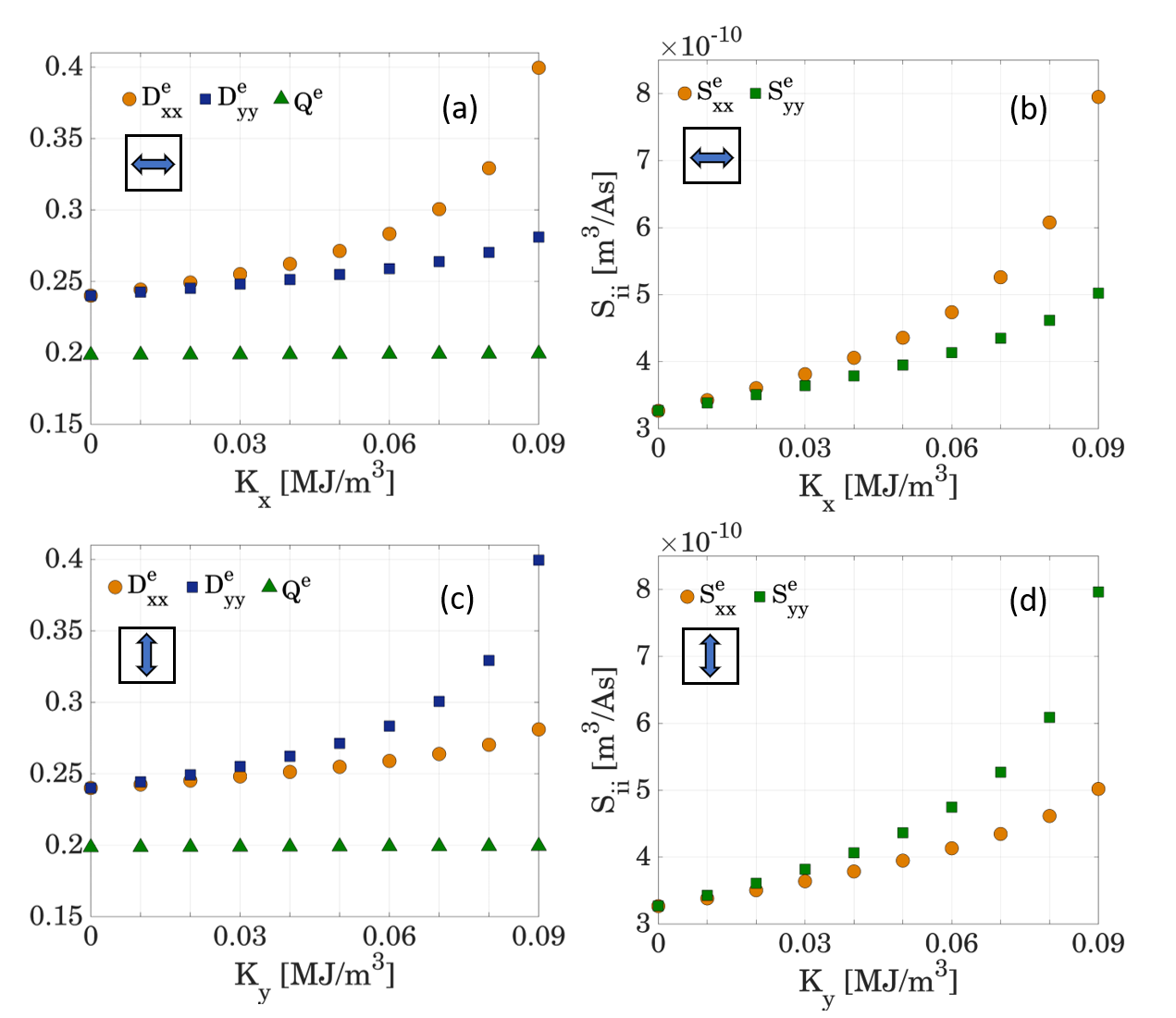}\caption{\label{fig:Figure3} (Colour online). The non-zero tensor elements of $\pmb{\mathsf{D}}^{e}$, $\pmb{\mathsf{S}}^{e}$ and the effective charge $Q^{e}$ as a function of induced in-plane anisotropies along $x$ and $y$ -directions, obtained from micromagnetically computed configurations. Parameters used are: $\gamma_{1}=\gamma_{2}=\text{2.21}\times\text{10}^{5}\text{m/As}$, $\phi_{1}=\text{0.1}$, $\phi_{2}=-\text{0.1}$, $o_{1}=\text{1}$, $o_{2}=-\text{1}$, $M_{1}=\text{0.6}$ MA/m, $M_{2}=\text{0.75}$ MA/m, $\alpha_{1}=\alpha_{2}=$ 0.1 and $d_{1}=d_{2}=$ 0.8 nm.  The direction of uniaxial in-plane anisotropy is indicated in each plot by double arrows. (a): $\mathsf{D}^{e}_{xx}$,$\mathsf{D}^{e}_{yy}$ and $Q^{e}$ vs. $K_{x}$. (b): $\mathsf{S}^{e}_{xx}$,$\mathsf{S}^{e}_{yy}$ vs. $K_{x}$. (c): $\mathsf{D}^{e}_{xx}$,$\mathsf{D}^{e}_{yy}$ and $Q^{e}$ vs. $K_{y}$ and (d): $\mathsf{S}^{e}_{xx}$,$\mathsf{S}^{e}_{yy}$ vs. $K_{y}$.}
\end{figure}

\begin{figure}[H]
	\includegraphics[width=\columnwidth]{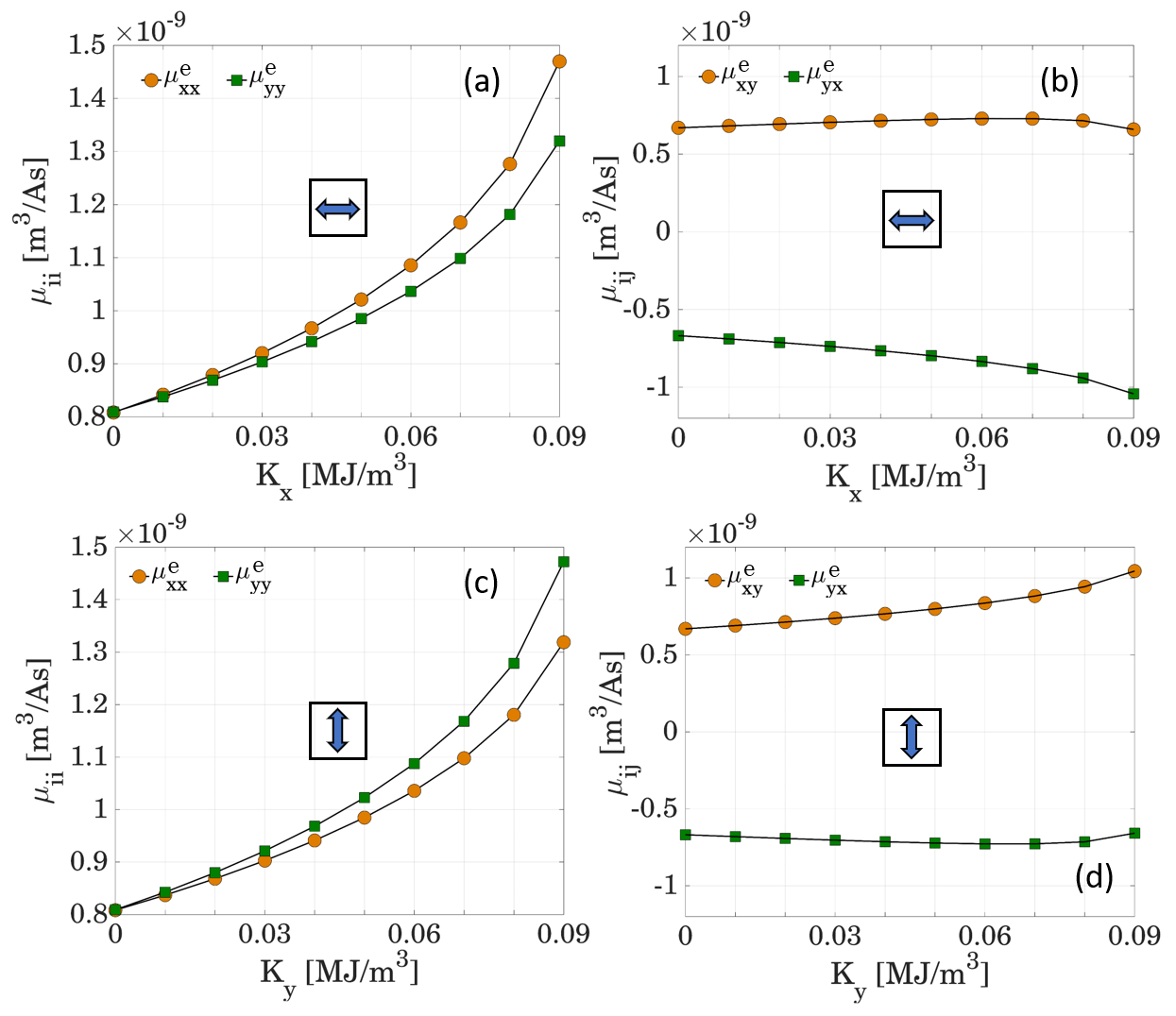}\caption{\label{fig:Figure4} (Colour online) Effective mobility-tensor elements as a function of induced in-plane anisotropy. Here, $k_{s}=$ 2. (a): $\mu^{e}_{xx}$ and $\mu^{e}_{yy}$ vs. $K_{x}$. (b): $\mu^{e}_{xy}$ and $\mu^{e}_{yx}$ vs. $K_{x}$. (c): $\mu^{e}_{xx}$ and $\mu^{e}_{yy}$ vs. $K_{y}$. (d): $\mu^{e}_{xy}$ and $\mu^{e}_{yx}$ vs. $K_{y}$.}
\end{figure}

\begin{figure} %[H]
\includegraphics[width=\columnwidth]{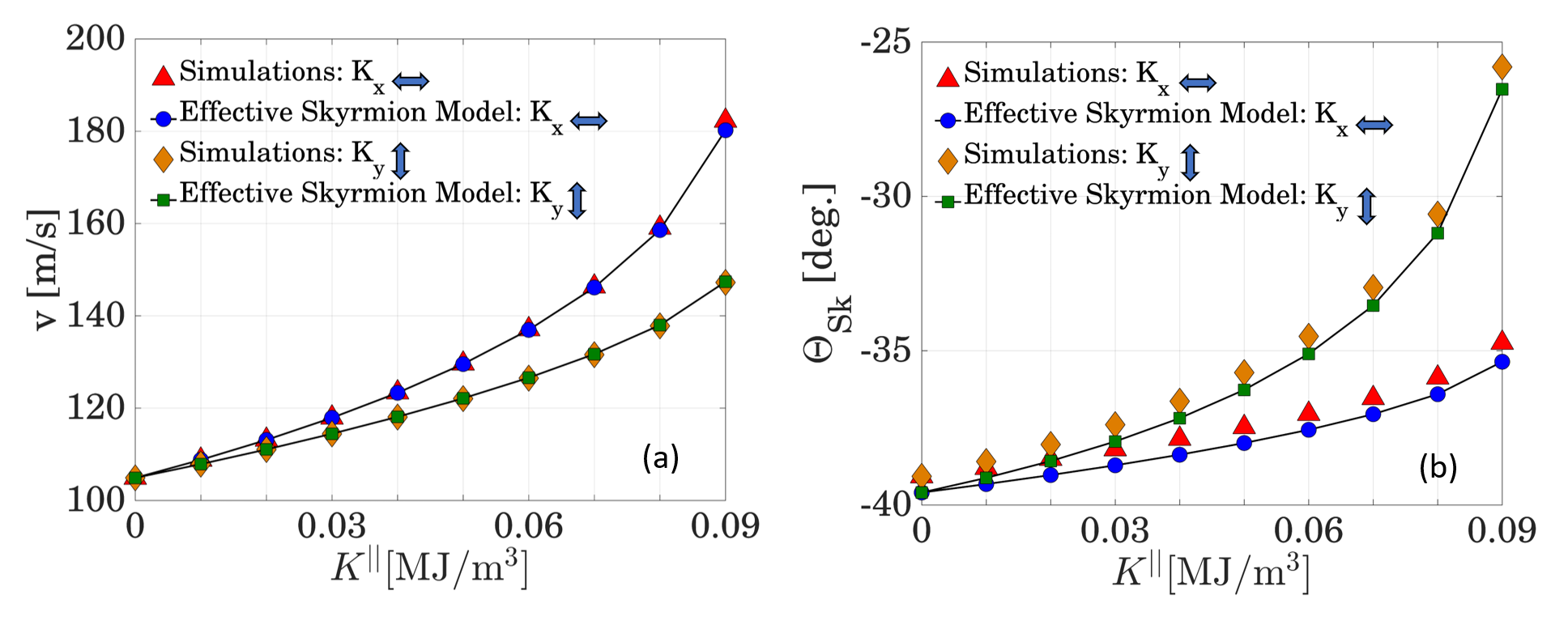}\caption{\label{fig:Figure5} (Colour online) speed, $v$ and $\Theta_{Sk}$ vs. $K_{x}$ with current injected along $x$ ($\theta_{J}=0$). The parameters used are the same as before. (a): $v$ vs. $K^{\parallel}$ ($K_{x}$ and $K_{y}$) predicted by the effective skyrmion approach and computed by full dynamic micromagnetic simulations. The direction of induced in-plane anisotropy is indicated by double arrows in the legends. (b): $\Theta_{Sk}$ vs. $K^{\parallel}$ ($K_{x}$ and $K_{y}$).}
\end{figure}

\begin{figure}[H]
	\includegraphics[width=\columnwidth]{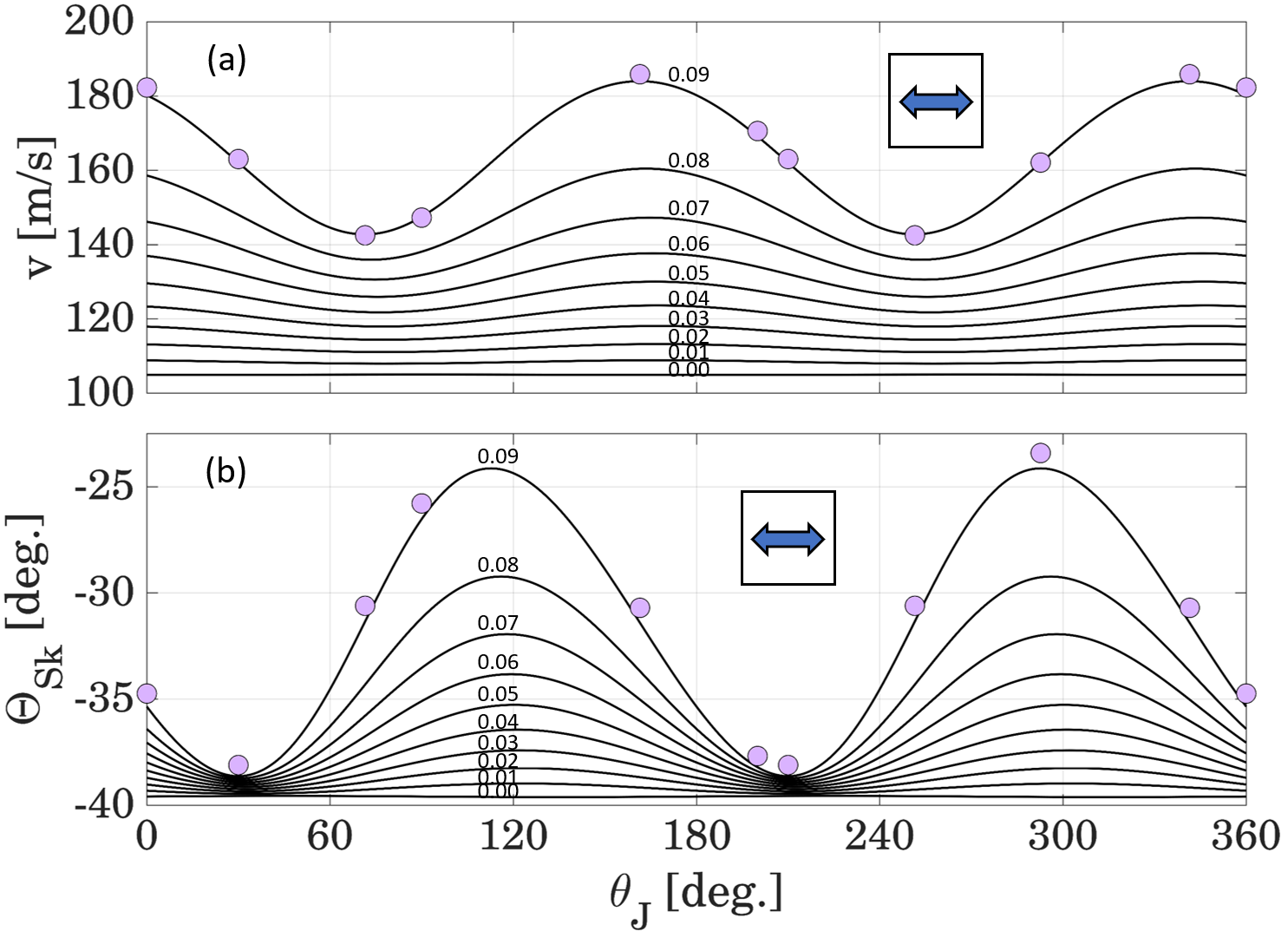}\caption{\label{fig:Figure6} Effective skyrmion prediction of (a): $v$ vs. $\theta_{J}$ for different $K_{x}$ and (b): $\Theta_{Sk}$ vs. $\theta_{J}$ for different $K_{x}$. Solid lines correspond to the effective skyrmion approach and filled circles are results from full dynamical micromagnetic simulations. The value of induced anisotropy $K_{x}$ (whose direction in the plane is also indicated by the double arrows) corresponding to each curve is shown by a number in units of MJ/$\text{m}^{3}$.}
\end{figure}

In order to quantify the effect of an in-plane anisotropy, equilibrium configurations were computed for a range of induced in-plane anisotropy states. A note however, is in order for our choice of the particular degree of ferrimagnetism, i.e. the choice of $M_{2}$ in relation to $M_{1}$. Firstly, it is found that for a given in-plane anisotropy constant, the larger the ratio of $M_{2}$ to $M_{1}$, the larger was the resulting skyrmion deformation. Secondly, although the stability of the bound skyrmion state ranges over a wide set of $M_{2}/M_{1}$-ratios in our system, $1\le M_{2}/M_{1}\le 1.5$, the range of applied in-plane anisotropy, whereby the bound skyrmion state is stable reduces as $M_{2}/M_{1}$ increases. We found $M_{2}/M_{1}=$ 1.25 to be a good compromise between the applicable range of imposed in-plane anisotropies and the degree of skyrmion deformation (which in turn means range of dynamical tuneability). A qualitative argument for the trend of increased skyrmion deformability with increasing $M_{2}/M_{1}$-ratios can be formed by considering that the dipolar interaction between $\text{FM}_{1}$ and $\text{FM}_{2}$ is frustrated, except in the region where the magnetization lies predominately in plane (within the skyrmion width). When a uniaxial anisotropy is imposed, the dipolar energy can be somewhat lowered by tilting more moments towards the plane. The projection on the plane of this tilting should of course be in the direction of the induced in-plane anisotropy. Thus, the higher the dipolar frustration is (meaning as the ratio $M_{2}/M_{1}$ increases), the higher is the motivation to increase the in-plane portion of the skyrmion state. This would translate to a higher degree of skyrmion deformation for a given value of the in-plane uniaxial anisotropy. Fig. \ref{fig:Figure2} shows the equilibrium configurations of the bound skyrmion state for a subset of induced in-plane anisotropies. The effect of a $K_{x}$ or $K_{y}$ can be stated by the following two observations: (i): The skyrmions elongate preferentially along the direction perpendicular to the induced anisotropy, i.e. become elliptical with the major axis perpendicular to the induced anisotropy direction. (ii): the over-all skyrmion-size increases with increasing in-plane anisotropy. Point (i) is a logical consequence of the system increasing its number of moments pointing by rotating the in-plane portion of the skyrmion towards the direction of induced anisotropy. Point (ii) is also to be expected, as by including an in-plane anisotropy, the out-of-plane anisotropy is in effect reduced. Test calculations were performed verifying that the skyrmion size increases as the perpendicular anisotropy decreases. Another contribution to skyrmion size-enhancement was touched upon above, in terms of lowering the dipolar frustration. If the texture was circularly symmetric with only the radius varying we should expect that the diagonal elements of the effective mobility tensor (Eq. \ref{eq:mobility}) to be equal and the absolute values of the off-diagonal elements to be equal, because then the diagonal elements of $\pmb{\mathsf{D}}^{e}$ are equal and the same would apply to $\pmb{\mathsf{S}}^{e}$, while $Q^{e}$ does not depend on the spin-profile of the skyrmion and thus remains constant [compare to circularly symmetric skyrmions in single layer FMs \cite{SkyHa1}]. In such a case, modulation of values of the tensor elements will indeed alter the speed and magnitude of $\Theta_{Sk}$, but the dynamics will be isotropic in the plane, i.e. no matter in which direction the driving current flows, the speed and $\Theta_{Sk}$ do not change. This could be achieved as has been proposed in other works; by modulating the perpendicular anisotropy constant \cite{modulation1}. The situation changes drastically if the skyrmion is deformed. Immediately we can expect that the elements of the diagonal tensors $\pmb{\mathsf{D}}^{e}$ and $\pmb{\mathsf{S}}^{e}$ differ in magnitude. This is shown in Fig. \ref{fig:Figure3}, whereby the dependence of $\pmb{\mathsf{D}}^{e}$, $Q^{e}$ and
$\pmb{\mathsf{S}}^{e}$, on induced in-plane anisotropies along $x$ and $y$ -directions are shown. The elements were computed by using the micromagnetically obtained configurations with parameters described in the caption. It was verified through direct calculations, that $\mathsf{D}^{e}_{xy}=\mathsf{D}^{e}_{yx}=0$ and $\mathsf{S}^{e}_{xy}=\mathsf{S}^{e}_{yx}=0$.
We can see the above discussion of the expectations verified, that for zero in-plane anisotropy, the elements of the diagonal tensors $\pmb{\mathsf{D}}^{e}$ and $\pmb{\mathsf{S}}^{e}$ are equal, signifying circular skyrmion shape. Let us look at Fig. \ref{fig:Figure3} (a) and (b). As $K_{x}$
increases there is a splitting in the dependencies of $\mathsf{D}^{e}_{xx}$ and $\mathsf{D}^{e}_{yy}$ in (a) and the same for $\mathsf{S}^{e}_{xx}$ and $\mathsf{S}^{e}_{yy}$ in (b). The charge $Q^{e}$ is as expected unaffected by the presence of the anisotropy-induced texture deformation. As $K_{x}$ increases, the skyrmions elongate more along $y$  (Fig. \ref{fig:Figure2} (b) and (c)). The effect on the drag force is then according to Fig. (\ref{fig:Figure3} (a)) a rapid increase along the direction of the elliptically shaped skyrmion's minor axis and a slower increase along the axis of elongation (major axis). If we think of the drag as a resistance to flow then we should expect a larger drag in the direction of propagation with the larger frontal area. Conversely along a direction whereby the object is more stream-lined-shaped, a relatively smaller drag is to be expected. This view is consistent with the observation made herein concerning sharp increase in $\mathsf{D}^{e}_{xx}$ as $K_{x}$ increases. The reason for a noticeable increase also in the $\mathsf{D}^{e}_{yy}$-element is that the over-all size of the skyrmion also increases with increasing $K_{x}$. In terms of $\mathsf{S}^{e}_{xx}$ and $\mathsf{S}^{e}_{yy}$ (Fig. \ref{fig:Figure3} (b)) which constitute the spin-Hall effect mobility, the largest increase is for $\mathsf{S}^{e}_{xx}$ as the number of moments along $x$ increases both as a result of satisfying the direction of anisotropy. The reasons for an increase of $\mathsf{S}^{e}_{yy}$ is similarly to the discussion on the drag force, due to an increase in skyrmion width, meaning an increase also in the number of moments along the $y$-direction. Since the SHE-induced spin-accummulation is perpendicular to the current direction, then we should expect that e.g. given a current fixed along $x$, the speed enhancement is greater for an induced anisotropy $K_{x}$ than for an induced $K_{y}$. The situation is reversed if the current is injected along $y$. The same arguments can be applied to the case of an induced anisotropy $K_{y}$ along $y$. This is shown in Figs. \ref{fig:Figure3} (c) and (d), whereby the situation is reversed with respect to Figs. \ref{fig:Figure3} (a) and (b), because the skyrmion preferential elongation is along $x$. We now evaluate the over-all mobility tensor using $\pmb{\mathsf{D}}^{e}$, $\pmb{\mathsf{S}}^{e}$ and $Q^{e}$. In order to be as general as possible we consider the case whereby $J_{1}\ne J_{2}$ and set here $k_{s}=$ 2. Evaluating the expression in Eq. (\ref{eq:mobility}) and plotting the tensor elements versus induced in-plane anisotropy along both $x$ and $y$ -directions, $K_{x}$ and $K_{y}$, respectively, we obtain the results shown in Fig. \ref{fig:Figure4}. 
As can be seen in Figs. \ref{fig:Figure4} (a) and (c), both mobilities $\mu^{e}_{xx}$ and $\mu^{e}_{yy}$ increase with increasing induced anisotropy. The rate of increase of these elements is greatest in the direction of the induced anisotropy. In terms of the off-diagonal elements, there is also a difference in their behaviour in terms of their rate of change with respect to the direction of induced anisotropy, see Figs. \ref{fig:Figure4} (b), (d). These two behaviours mean that the speed $v$ and skyrmion Hall-angle $\Theta_{Sk}$ are tuneable by the in-plane anisotropy and that we can have anisotropy in this tuneability by two means; either by, for a given current direction, induce the anisotropy along different directions, or for a fixed induced anisotropy-direction, vary the angle of the driving current. In what manner $v$ and $\Theta_{Sk}$ changes with increased magnitude of the induced anisotropy boils down to the relative rate of change and magnitudes of the tensor elements in Fig. \ref{fig:Figure3} as the induced anisotropy is varied. The speed for a given induced anisotropy gets its largest contribution from the diagonal elements of $\pmb{\mu}^{e}$. In conjunction with the relative rate of growth of these elements being large we can expect that for a given fixed current-direction, $v$ will always increase with increasing magnitude of in-plane anisotropy. In addition, it will become clear that the source of effective skyrmion deflection away from the driving current direction can in principle stem from two sources (except for current-directions exactly along $x$ or $y$ -directions): one due to a finite topological charge (the dominant contribution) and the second comes purely from lateral shape-distortions of the skyrmions. In fact, even for a situation whereby $Q^{e}$=0 such as in a perfectly balanced SAF, there can be a deflection away from the driving current-angle for $\theta_{J}$ different from multiples of $\pi/2$. We shall now address these issues in an orderly fashion. Although lengthy, it is instructive to write out the full expressions for $\Theta_{Sk}$ and $v$ given an arbitrary current direction. Recalling that $\mathbf{J}=J[\cos\left(\theta_{J}\right)\text{ } \sin\left(\theta_{J}\right)]^{T}$ with $\Theta_{Sk}$ defined with respect to the primed coordinate system and $v$ being a magnitude which is easily stated using the global system as starting point (as the magnitude will be the same in both coordinate systems), one arrives at:
\begin{widetext}
\begin{align}
v&=J\sqrt{\sin\left(2\theta_{J}\right)\left[\mu_{xx}^{e}\mu_{xy}^{e}+\mu_{yx}^{e}\mu_{yy}^{e}\right]+  \cos^{2}\left(\theta_{J}\right)\left[\left(\mu_{xx}^{e}\right)^{2}+\left(\mu_{yx}^{e}\right)^{2}\right]+\sin^{2}\left(\theta_{J}\right)\left[\left(\mu_{xy}^{e}\right)^{2}+\left(\mu_{yy}^{e}\right)^{2}\right]}\label{eq:full_speed}\\[4pt]
\Theta_{Sk}&=\arctan\Bigg\lbrace\frac{\frac{1}{2}\sin\left(2\theta_{J}\right)\left[\mu_{yy}^{e}-\mu_{xx}^{e}\right]+\mu_{yx}^{e}\cos^{2}\left(\theta_{J}\right)-\mu_{xy}^{e}\sin^{2}\left(\theta_{J}\right)}{\frac{1}{2}\sin\left(2\theta_{J}\right)\left[\mu_{xy}^{e}+\mu_{yx}^{e}\right]+\mu_{xx}^{e}\cos^{2}\left(\theta_{J}\right)+\mu_{yy}^{e}\sin^{2}\left(\theta_{J}\right)}\Bigg\rbrace\label{eq:full_angle}
\end{align}
\end{widetext}
Let us start with tuneability of $v$ and $\Theta_{Sk}$ by induced $K_{x}$ and $K_{y}$ for a fixed $\theta_{J}$: Consider, as an example the case $\theta_{J}=$0 (i.e. current injected purely along the $x$-direction). Then, from Eq. (\ref{eq:full_speed}) and Eq. (\ref{eq:mobility}), $v\left(\theta_{J}=0\right)=J\sqrt{(\mu_{xx}^{e})^{2}+(\mu_{yx}^{e})^{2}}=\frac{JS_{xx}^{e}}{\left(Q^{e}\right)^{2}+D_{xx}^{e}D_{yy}^{e}}\sqrt{\left(D_{yy}\right)^{2}+\left(Q^{e}\right)^{2}}$. From this expression, it is then clear that the speed will have a different dependency on $K_{x}$ (with $K_{y}=0$) than on $K_{y}$ (with $K_{x}=0$), based on the individual behaviour of the individual tensor elements on the induced anisotropy. Now, for $\Theta_{Sk}$, then from Eq. (\ref{eq:full_angle}) and Eq. (\ref{eq:mobility}), $\Theta_{Sk}\left(\theta_{J}=0\right)=\arctan\left[\frac{v_{y'}\left(\theta_{J}=0\right)}{v_{x'}\left(\theta_{J}=0\right)}\right]=\arctan\left[\frac{\mu_{yx}^{e}}{\mu_{xx}^{e}}\right]=\arctan\left[\frac{-Q^{e}}{D_{yy}^{e}}\right]$, which is the standard result for the skyrmion-Hall angle [Ref] as a special case of $\theta_{J}=0$, except that we specifically state here the $D_{yy}^{e}$-element in the denominator. We can intuitively predict the behaviour of $\Theta_{Sk}$. Let us consider again $\theta_{J}=$ 0 (current flowing along $x$) and an induced anisotropy $K_{x}$; We know from Fig. \ref{fig:Figure3}, that with increasing $K_{x}$, the largest change in dissipation occurs for the $D_{xx}^{e}$ element, but $D_{yy}^{e}$ increases as well with the result that the magnitude of $\Theta_{Sk}$ is expected to decrease at some rate. Now, keeping the same scenario except now we impose an anisotropy $K_{y}$, we know that the rate of increase of $D_{yy}^{e}$ is even greater with increasing anisotropy $K_{y}$. The expectation then of the change in magnitude of $\Theta_{Sk}$ should be expected to be greater for induced anisotropy along $y$ when driving with $\theta_{J}=$ 0. To illustrate this simple analysis, $v$ and $\Theta_{e}$  as a function of $K_{x}$ and $K_{y}$ for the current-angle $\theta_{J}=0$ is computed and shown in Fig. \ref{fig:Figure5}. We have also compared the effective skyrmion approach to full dynamical micromagnetic simulations with an excellent agreement, thus validating the approach. The overall range of tuneability is significant and is anisotropic with respect to induced-anisotropy direction. In a real situation we could envision the device mounted on a piezoelectric stressor that can transmit strain in two ortogonal directions (one at a time) in order to induce and discriminate different speeds.  
We now address the general case $\theta_{J}$ is allowed to vary whereby anisotropic behaviour of speed and deflection angle are expected to be present. In addition, we point out, that in general, as long as $\theta_{J}\ne n\pi/2$ (where $n$ is an integer) and the effective skyrmion radius is not isotropic (deviation from circular shape) , $\Theta_{Sk}$ will contain a contribution originating from only the anisotropic deformation of the skyrmion, i.e. independent of the topological charge. In addition the speed may also be non-isotropic with respect to driving-current direction. In order to see this, we briefly digress on this point as it may be of consequence for experiments whereby there is a sizeable inverse magnetostrictive effect causing a deviation from circularly shaped skyrmions. If we impose $Q^{e}=0$ (in practice this means balancing the FM layers such that $w=$ 1), such that all off-diagonal elements in Eq. (\ref{eq:mobility}) are zero, then From Eqs. (\ref{eq:full_speed}) and (\ref{eq:full_angle}),  $v\left(Q^{e}=0\right)=J\sqrt{\left(\mu_{xx}^{e}\right)^2\cos^{2}\left(\theta_{J}\right)+\left(\mu_{yy}^{e}\right)^2\sin^{2}\left(\theta_{J}\right)}$ and $\Theta_{Sk}\left(Q^{e}=0\right)=\arctan\left[\frac{\frac{1}{2}\sin\left(2\theta_{J}\right)\left[\mu_{yy}^{e}-\mu_{xx}^{e}\right]}{\left(\mu_{xx}^{e}\right)^2\cos^{2}\left(\theta_{J}\right)+\left(\mu_{yy}^{e}\right)^2\sin^{2}\left(\theta_{J}\right)}\right]$. It is now clear that, provided $\mu_{xx}^{e}\ne\mu_{yy}^{e}$, there will be a finite $\Theta_{Sk}$ for all current-angles $\theta_{J}\ne n\frac{\pi}{2}$ in the absence of a net topological charge. In terms of the speed, we can also see that the role of a topological charge on $v$ is two-fold; it affects both magnitude and shifting the value of $\theta_{J}$ whereby $v$ has its maximum or minimum; with $Q^{e}=0$, $v$ exhibits maxima and minima at $\theta_{J}=n\frac{\pi}{2}$, but for $Q^{e}\ne 0$, the $\sin\left(2\theta_{J}\right)$ term in Eq. (\ref{eq:full_speed}) imposes a shift away from $n\frac{\pi}{2}$. We now return to our original easily deformable system and make predictions for a case whereby for fixed values of $K_{x}$ (with $K_{y}=0$), we sweep the injected current angle $\theta_{J}$. The results for the dependence of both $v$ and $\Theta_{Sk}$ are shown in Fig. \ref{fig:Figure6}. Results are also compared for the highest value of $K_{x}$ to full dynamic micromagnetic simulations as a verification-step of the effective skyrmion prediction, with a very good match. The amplitudes of oscillations in the dynamical behaviours are quite significant and should be easily detectable in an experiment. Apart from the additional degree of freedom in terms of modulating the dynamics, this anisotropic behaviour could be used to detect the possible presence of strain induced anisotropy in the system and thus of skyrmion deformation. 
\section{Conclusions}
In conclusion, we have shown, by combined micromagnetic simulations and an effective skyrmion analytical model, that we can effectively modulate both speed and skyrmion Hall-angle of tightly antiferromagnetically bound skyrmions by induced in-plane anisotropies. The cause of the said modulations stem from a deformation of the skyrmion-texture (going from circular to elliptical shapes). As a further consequence, we showed that this introduces dynamical anisotropy in the plane of skyrmion propagation with respect to driving-current injection-angle. In addition, we have shown, given a deviation from circularly shaped skyrmions, that for driving current angles $\theta_{J}\ne n\frac{\pi}{2}$, there is a contribution to the skyrmion deflection away from the driving-current direction independent of the topological charge, i.e. even for a perfectly balanced SAF. This may be of consequence for SAF devices whereby skyrmions operate on relatively large areas, causing a build-up of in time of deviation from the intended target position. Finally, if the uniaxial anisotropies can be induced by mechanical stress, it can possibly lead to less complex device structures as compared to other proposed schemes.

\end{document}